\documentstyle[prl,twocolumn,aps]{revtex}

\begin{document}

\twocolumn[\hsize\textwidth\columnwidth\hsize\csname @twocolumnfalse\endcsname

\draft

\title{ {\bf 
Theory of extraordinary optical transmission through
subwavelength hole arrays }} 

\author{L. Mart\'{\i}n-Moreno$^1$, F. J. Garc\'{\i}a-Vidal $^2$,
H. J. Lezec$^3$, K.M. Pellerin $^4$, T. Thio $^4$,
 J. B. Pendry $^5$, T. W. Ebbesen $^3$ \\} 

\address{
$^1$ {\it Departamento de F\'{\i}sica de la Materia Condensada,
ICMA-CSIC,
Universidad de Zaragoza, E-50015 Zaragoza, Spain.} \\
$^2${\it Departamento de F{\'{\i}}sica Te\'orica
de la Materia Condensada,
 Universidad Aut\'onoma de Madrid,
 E-28049 Madrid, Spain} \\
$^3$ {\it ISIS, Universit\'e Louis Pasteur, 67000 Strasbourg, France}
\\
$^4$ {\it NEC Research Institute, Princeton, New Yersey 08540, USA} \\
$^5$ {\it The Blackett Laboratory, Imperial College, London SW7 2BZ, United
Kingdom} \\
}
\maketitle
\begin{abstract}
We present a fully three-dimensional theoretical study of the 
extraordinary transmission of light through  subwavelength hole arrays 
in optically thick metal films. Good agreement is obtained with
experimental data. An analytical minimal model is also developed, which
conclusively shows that the enhancement of transmission
is due to tunneling through 
surface plasmons formed on each metal-dielectric interfaces. 
Different regimes of tunneling (resonant through a ''surface plasmon
molecule", or sequential through two isolated surface plasmons) 
are found depending on the geometrical parameters 
defining the system.

\end{abstract}

\pacs{PACS numbers: 78.66.Bz, 73.20.Mf, 42.79.Dj, 71.36.+c }

]

\narrowtext

In the last few years, and mainly due to advances in nanotechnology,
there has been a renewed interest in exploiting the dielectric response of
metals to make photonic materials\cite{Pendry99,Fan96,Sievenpiper98}.
For instance, the photonic
insulating properties of metals can be used to trap incident radiation,
focusing light in very small 
volumes\cite{Barnes96,Garcia96,Krenn99}. 
Very recently\cite{Ebbesen}, another interesting effect of light 
interacting with structured metals has been discovered:
the transmission of light through subwavelength hole arrays made
in a metal film can be orders of magnitude larger than 
expected from standard aperture theory\cite{Bethe44}.
Apart from its fundamental interest,
this extraordinary transmission effect has  
potential applications\cite{Sambles98,Thio00} in subwavelength 
photolithography, near-field
microscopy, wavelength-tunable filters, optical modulators, and flat-panel
displays, amongst others. While the wavelength at which some transmission
features appeared suggested\cite{Ebbesen,Grupp00} that 
surface plasmons (SP)\cite{Rather88} were involved in the
process, the physical mechanism for the huge 
enhancement has not yet been
elucidated. Some calculations have been performed for
a simpler geometry: an array of 
slits\cite{Schroter98,Treacy99,Porto99}, where high transmission was also
predicted.
However, although interesting in their own right, these results do not apply
to the experimental situation. 

In this letter, we present the first fully three-dimensional
theoretical study of transmission through hole arrays, obtaining an excellent
agreement with experimental data. Moreover, we develop a
simplified version of the model that clearly captures the physics involved.

Fig. 1 shows the experimental
``zero-order" transmittance of light 
at normal incidence ($T_{00}$),
through an
array of holes in a free-standing metal film.  The free-standing metal
film, of which the fabrication is described elsewhere \cite{Grupp00} consisted
of a $220nm$ thick Ni core, perforated with a square array of holes by
focused-ion beam milling.  The film was subsequently overcoated with
$50nm$ of Ag on both sides by sputter deposition which resulted in a coating of
the walls of the holes as well as the in-plane surfaces of the film.
The total thickness of the film was $h=320nm$ and the lattice constant
of the hole array was $L=750nm$.  After
coating the holes had an average diameter of $280nm$.
It has been shown \cite{Grupp00}
that such a "sandwich" structure has the same transmission properties as
an equivalent perforated film made of silver throughout.
The advantage of a free-standing metal film is that it is possible to
mill much better defined holes than in a film on a substrate.  Moreover, in a
free-standing film the dielectric constant is the same in all
non-metallic regions.
In Fig. 1, $T_{00}$ shows the well known 
Rayleigh minima of Wood's anomaly \cite{Wood}
which appear in any diffractive array roughly when an order of
diffraction emerges
tangent to the array. The data also shows the extraordinary
transmittance effect: at $\lambda \approx  800 nm$, $T_{00}$  
of the order of 15\% was found. As the area
covered by holes is only 11\%, the normalised-to-area transmittance of
light is 130\%. Standard aperture theory for a single hole predicts
transmission efficiencies of order 1\%.

In our theoretical model, we consider a metal film of thickness $h$,
perforated with square holes of side length $d$, periodically distributed,
as in the experiment, in a square array of lattice parameter $L$. 
In the long-wavelength limit, experimental results have shown that the
transmission coefficient depends on hole area, but does not appreciably 
depend on hole shape. 
The choice of square holes is, therefore, inessential, 
and was motivated purely for analytical convenience. When comparing with
experimental results we take the square side as the square root of
experimental hole area. We consider the general case in which light impinges
from a uniform medium 1, into the perforated metal film (region 2) 
which lies on a substrate (medium 3),
where the light in collected. The dielectric constants are, $\epsilon_1$ in
medium 1, $\epsilon_2$ inside the holes, and $\epsilon_3$ in medium 3. The
metal is characterised by a frequency-dependent complex dielectric constant $%
\epsilon_m(\omega)$.

We treat the EM fields in the metal through the surface impedance boundary
condition\cite{Jackson75}, a procedure fully justified, as the frequencies
considered are well below the metal
plasma frequency. We expand the EM fields
in the eigenmodes of the different regions. That is, the EM fields are a
linear expansion of S and P plane waves in regions 1 and 3, and Bloch waves
combining (evanescent and, if there is any, propagating) 
$TE$ and $TM$ waveguide modes\cite{Jackson75} inside the holes. 
Although, in principle, an infinite number of modes in each region should 
be taken into account,
results quickly converge with the number of parallel wavevector components
considered. The probability amplitude, $t_{if}$, for an arbitrary incident
(from medium 1) EM plane wave, $i$, to be transmitted to a outgoing plane
wave $f$, in medium 3, was calculated in a multiple scattering formalism. By
appropriately matching the EM fields in the 1-2 and 2-3 interfaces, $t_{if}$
can be expressed in terms of transmission and reflection amplitudes for a
single interface as:

\begin{equation}
t_{if} \, = \, \sum_{\alpha,\beta,\gamma}\,\tau^{12}_{i\alpha} \, e_{\alpha}
\left( \delta_{\alpha \beta} \, - \, \rho^{R}_{\alpha \gamma} \,
e_{\gamma}\, \rho^{L}_{\gamma \beta} \, e_{\beta} \right) ^{-1} \,
\tau^{23}_{\beta f}
\end{equation}

Latin indexes refer to plane waves, either in medium 1 or 3, and Greek
indexes to waveguide modes inside the holes. $\tau^{12}, \tau^{23}, \rho^{R} 
$ and $\rho^{L}$ are components of the scattering matrix for a
single interface, either 1-2 or 2-3. More precisely, $\tau^{12}$ is a matrix
giving the transmission amplitude between modes in medium 1 to modes in
medium 2. $\rho^{R}_{\alpha \gamma}$ is the reflection amplitude for the
waveguide mode $\alpha$, traveling towards medium 1, to be reflected into
the mode $\gamma$ traveling away from medium 1, after scattering with the
2-1 interface. $\rho^{L}_{\alpha \gamma}$ denotes the same but for the
interface 2-3. $\tau^{23}_{\beta f}$ connects mode $\beta $ with outgoing
mode $f$, through interface 2-3. Finally, $e_{\alpha} = e^{\imath
q_{z\alpha} h}$, $z$ is the direction perpendicular to the interfaces, and $%
q_{z\alpha}$ is the wavevector component along the z-direction for the
waveguide mode $\alpha$.

Figure 2 shows the computed results for $T_{00}$.
Calculations were done for the nominal
experimental values for the different parameters defining the structure.
The metal dielectric constant has been taken as the measured
one for Ag\cite{epsilonAg}. It must be stressed that in our calculation
there are no fitting parameters. Also shown are results for slightly
different values for hole size, $d$, in order to give an idea of the
sensitivity of the results to non-uniformities of hole area. As shown in 
Fig. 2, our calculation also shows peaks,
with $T_{00}$ which exceeds by 2-3 orders of magnitude the
predictions by a theory
based on independent holes\cite{Bethe44}. Moreover, the wavelenghts at which
peaks occur closely match the experimental values. Clearly the 
mechanism responsible for the extraordinary transmission is present in our
model. Discrepancies between calculated and experimental values for
linewidths and maximum transmittances, can well be due to small variations
in the hole diameter throughout the sample, as supported by the
sensitivity of the transmittance to $d$ shown in Fig. 2.

Although the calculation also shows extraordinary transmission through
periodic subwavelength holes in a metal, it is not obvious what 
mechanism is involved. However, as we show in the following, a highly
simplified version of the model,
which can be analytically worked out,
also shows the same behaviour in the regime of interest: $
\lambda \ge L \gg d$. In this minimal model, a strong truncation in parallel
wavevectors is made: in regions 1 and 3, only first order diffraction
is considered, i.e., the possible wavevectors in the direction of the
incoming electric field, $x$, are $k_{x0}=0$ and $k_{x\pm 1} =
\pm \frac{2 \pi}{L}$; in region 2,
where all modes are evanescent for $L>2d$, only the most slowly decaying
evanescent mode, the $TE_{01}$ mode, is taken into account. Figure 3 shows the
comparison of the fully converged calculation (solid line) with the results
obtained with the minimal model (dashed line), for the nominal parameters of
the experimental set up, both for (3a) the experimental dielectric constant
for Ag and (3b) for an hypothetical case of non-absorbing Ag 
($ Im\left[\epsilon(\omega)\right]=0$). The minimal model captures 
well the extraordinary
transmission phenomena: the main effect of truncating the number of modes is
a small unimportant shift in the position of the transmission peaks.
In order to further simplify the discussion, in the following we 
consider the ideal non-absorbing case. 
Finite absorption merely reduces the height of the
peaks (for zero absorption the transmittance can be as large as 100 \%)
without altering the physical picture. 

For the symmetric case, 
$\epsilon_1 = \epsilon_2=\epsilon_3=1$, 
$\rho^R_{11} = \rho^L_{11}\equiv\rho$ and 
$t_{00}$ simplifies to

\begin{equation}
t_{00} \, = \, \frac{\tau^{12}_{01}\,e^{-|q_{z1}|h} \tau^{23}_{10}}{ 1 \, -
\, \rho^2 \, e^{-2|q_{z1}|h}}
\end{equation}

where $\tau^{12}_{01} = 2 S_0/(G_2 + G_1),\,\, \tau^{23}_{10} = 2
G_2/(G_2 + G_1)$ and $\rho = (G_2 - G_1)(G_2 + G_1)$.
$G_2$ and $G_1$ are ``conductances" (inverse of total impedances) 
in region 2 and in region 1, 
respectively. They have somewhat involved analytical 
expressions that, for the case
of a hole array in a perfect metal, simplify to 
$G_2 = q_{z1}/g , \, G_1 = S_0^2 + 2 S_1^2 g/k_{z1}$, 
with $g=\omega/c, S_0=d/L,
S_1 = S_0 \sin\left[k_{x1} d/2\right]/(k_{x1} d/2)$, and 
$k_{z1} = \sqrt{g^2-k_{x1}^2}$.
 
Wood's anomaly is present in this model,  
associated to a zero in either $\tau^{12}_{01}$ or
$\tau^{23}_{10}$, i.e., when the conductance in medium 1 or 3, respectively,
is infinite.  

With respect to the extraordinary transmission, 
transmittance peaks $\approx$ 1 seem to point to the presence of resonant
phenomena. This idea is reinforced by the presence of multiple scattering
denominators in Eq.(2) that could be close to zero. Naively, one would expect
that this possibility never occurs: $e^{-2|q_{z1}|h} \ll 1$ in the regime
considered and $\rho$ is the amplitude for reflection, which can be expected
to be always less than 1. Actually, this is not true, as the condition $
|\rho | \le 1$ applies only for propagating modes, where it corresponds to
current conservation. In the regime we are considering, all modes inside the
hole are evanescent and current conservation only restricts 
$Im\left[(\rho\right] \ge 0$,
with no restrictions on the real part or the modulus of $\rho$
. This opens up the possibility of resonant denominators for $|\rho | \gg 1$
. In fact, it can be shown analytically that $|\rho |$ has a peak at
$\lambda$ slightly larger than $L$, with a maximum value that scales as $(
\frac{L}{d})^3$. Figure 4a shows $|\rho |$ as a function of $\lambda$ for
the parameters considered in Fig. 2. 
It is illuminating at this point to consider the dependence of the
transmittance on $h$. Figure 4b illustrates graphically 
that the peaks in $T_{00}$ 
occur at the $\lambda$ for which the distance between $|\rho|$
and $e^{|q_{z1}| h}$ is minimal. 
For zero absorption, a $T_{00} $ close to unity occurs whenever
this distance is zero, 
i.e. when the curves intersect. For large enough $h$, the
curves do not intersect and $T_{00}$ maxima 
decay exponentially with $h$.

All this is mathematics, showing that the effect of extraordinary
transmission has a resonant nature. But, what is the physical origin?
Analytically, we find that the frequency at which the maximum of $|\rho|$
appears coincides with the SP frequency of a
periodically  (air-metal with holes) isolated interface at parallel
momentum $\frac{2 \pi}{L}$ (what we are going to call first-order SP). This
is a localised EM mode, confined to the interface. It is not the only EM
mode at that frequency, as there also exists the zero-parallel-momentum
propagating mode, into which the SP can decay. This decaying channel
provides a linewidth and a maximum value to $\rho(\lambda$), which otherwise
would be a delta function. Furthermore, when we calculate the SP of the
periodically perforated metal slab, we find that
the single interface first-order SPs on the two metal surfaces
combine to form a ''SP molecule", in
much the same way as electronic states of isolated atoms combine to form
molecular levels. The frequencies of the first-order SP molecule are
given precisely by the same condition for resonant denominators
in the expression for $t_{00}$, if the coupling to the 
radiative modes is neglected.
As in the case of the isolated interface, this SP molecule coexists
with a continuum and, rather than an exact eigenstate, is a (more or
less long lived) resonance.

This point on the formation of the SP molecule can be more precisely
discussed by considering typical times in the transmission
process. The width of the peak of $\rho(\omega)$ is related to the lifetime
of the SP, $t_{rad}$, the time taken for a SP of an isolated
surface to decay into the radiative mode. We obtain that, approximately, 
$t_{rad}\approx (c |k_{z1}|)^{-1}$\cite{caution}. From the frequency splitting
between molecule levels, taken as uncoupled from the radiation modes, the time
needed for each of the molecular modes to form, $t_{res}$, can be
estimated\cite{caution} 
as $t_{res} \approx t_{rad} (d^3 e^{|q_{z1}| h}/L^3)$.
So, in the small $h$ regime, 
where the intersection of $|\rho|$ with the
exponential occurs at the bottom of the $|\rho|$ peak, 
$t_{res}\ll t_{rad}$ and the
molecule levels are fully formed. The photon then goes back and forth several
times inside the hole, building up coherent constructive
interference in the forward direction
much as would occur in electron resonant tunneling or in a
Fabry-Perot interferometer (but with evanescent waves instead of
propagating ones). As we consider larger $h$, $t_{res}$ increases, while $%
t_{rad}$, being a property of an isolated interface, essentially 
remains constant. So, the photon can make fewer round trips inside 
the hole before being radiated
to infinity, and the concept of plasmon molecule becomes less
well-defined. The
condition $t_{res}=t_{rad}$ marks the value of $h$ at which the
transmittance maximum ceases to be one. For even larger $h$, 
$t_{res} \gg t_{rad}$ and the
process is more like sequential tunneling, where the incoming photon gets
trapped in a SP, tunnels to the SP at the other interface and then couples
to the outgoing radiative mode and exits. The
transmittance is enhanced in this
case, as the photon can use two intermediate states (the first-order SP at both
interfaces) to cross the metal film, but the enhancement is not as efficient
as when the hopping occurs back and forth several times
building up a constructive interference.

Absorption introduces another time into the problem: $t_{abs}$, the typical
time it takes for a photon to get absorbed. Provided this time is smaller
than both $t_{res}$, and $t_{rad}$, the physical picture in not greatly
altered. This is the case for the experimental parameters:
absorption reduces the $T_{00}$ maxima by a factor around 2-3, without
altering either the position or the width of the peaks.

The above discussion is for the special case of matching dielectrics
($\epsilon_1 = \epsilon_3$) but the
picture still applies in the non-symmetric case ($\epsilon_1 \ne
\epsilon_3$), 
the situation in which extraordinary transmittance was
initially discovered. Here, SP on the two different interfaces of the
metal film have different frequencies and, instead of a SP diatomic
homopolar molecule forming in the presence of a continuum, 
we have instead a diatomic heteropolar molecule. From our model we find 
that there still are resonant peaks, but the transmittance maxima is reduced
with respect to the symmetrical case. 
A detailed comparison between experimental and theoretical
results in this case will be presented elsewhere.

In summary, we have presented a detailed theoretical study of the
extraordinary transmission of light through subwavelength hole
arrays in metal films, obtaining good agreement with experimental data on
free-standing films. An analytical minimal model, based on the
numerical calculations, conclusively shows that the 
the enhanced transmission is due to tunneling through surface plasmons. 
Different regimes in the enhanced transmission are found: 
for small film thickness the tunneling is resonant 
through ''plasmon molecule" levels while, for large thicknesses, 
the photon hops from surface plasmon to surface plasmon, but 
exits the structure before
the molecule level is developed, the tunneling being then sequential.
Undoubtedly, this found ability of surface plasmons to transmit and focus
light very efficiently is not restricted to this particular geometry and
will be exploited for controlling light in new photonic devices.

\begin{figure}[h]
\caption{Experimental zero-order transmittance of a square array of holes
(lattice constant $L = 750 nm$, average hole diameter of $280 nm$) 
in a free-standing
Ag film (thickness $h = 320 nm$). Inset: electron micrograph of
the perforated metal film.}
\end{figure}

\begin{figure}[h]
\caption{Calculated $T_{00}$ for 
an array of holes in
a Ag film, defined by $L = 750 nm,\, h = 320 nm$ and
three different hole side lengths $d$.} 
\end{figure}

\begin{figure}[h]
\caption{Comparison between the calculated $T_{00}$ within the full
(solid line) and minimal (dashed line) models, 
for the system considered in Fig.2,
with $d = 280 nm$. 
In panel (a) the dielectric constant is that of Ag, whereas in (b)
absorption of Ag is neglected.}
\end{figure}

\begin{figure}[h]
\caption{(a) $|\rho|$ versus $\lambda$ for the parameters 
defined in Fig. 2, within the minimal model with no-absorption 
(solid line). Also $e^{|q_{z1}| h}$ for different values of $h$ (in nm). 
(b) $T_{00}$ versus $\lambda$ for the  
metal thicknesses considered in panel (a) (successive curves being 
shifted by 100\%). 
}
\end{figure}


\begin{references}

\bibitem{Pendry99}
J. B. Pendry, Science {\bf 285}, 1687 (1999).

\bibitem{Fan96}
S. Fan, P. R. Villeneuve and J. D. Joannopoulos, Phys. Rev. B, {\bf 54}, 11245
(1996).

\bibitem{Sievenpiper98} 
D. F. Sievenpiper {\it et al.}, Phys. Rev. Lett. {\bf 80}, 2829 (1998).

\bibitem{Barnes96} 
W.L. Barnes {\it et al.}, Phys. Rev. B {\bf 51}, 11164 (1995);
W. L. Barnes {\it et al.}, Phys. Rev. B {\bf 54}, 6227 (1996).

\bibitem{Garcia96}
F. J. Garc\'{\i}a-Vidal and J. B. Pendry, Phys. Rev. Lett.
{\bf 77}, 1163 (1996).

\bibitem{Krenn99}
J. R. Krenn {\it et al.}, Phys. Rev. Lett. {\bf 82}, 2590 (1999).

\bibitem{Ebbesen}
T. W. Ebbesen {\it et al.}, Nature (London) {\bf 391}, 667 (1998);
H. F. Ghaemi {\it et al.}, Phys. Rev. B {\bf 58}, 6779 (1998);
T. J. Kim {\it et al.}, Opt. Letts. {\bf 24}, 256 (1999);
D. E. Grupp {\it et al.}, Adv. Mater. {\bf 11}, 860 (1999).

\bibitem{Grupp00}
D.E. Grupp {\it et al.}, Appl. Phys. Lett., in press (9/2000).

\bibitem{Bethe44}
H. A. Bethe, Phys. Rev. {\bf 66}, 163 (1944).

\bibitem{Sambles98}
J.R. Sambles, Nature (London) {\bf 391}, 641  (1998).

\bibitem{Thio00}
T. Thio {\it et al.}, Physica B {\bf 279}, 90 (2000).

\bibitem{Rather88}
H. R\"ather, {\it Surface Plasmons on Smooth and Rough Surfaces and on Gratings}
(Springer-Verlag, Berlin, 1988).

\bibitem{Schroter98}
U. Schroter and D. Heitmann, Phys. Rev. B  {\bf 58}, 15419 (1998).

\bibitem{Treacy99}
M. M. J. Treacy, Appl. Phys. Lett. {\bf 75}, 606 (1999).

\bibitem{Porto99}
J. A. Porto, F. J. Garc\'{\i}a-Vidal and J. B. Pendry, Phys. Rev. Lett.
{\bf 83}, 284 (1999).

\bibitem{Wood}
Lord Rayleigh, Proc. Roy. Soc. London Ser. A {\bf 79}, 399 (1907);
Philos Mag. {\bf 14}, 60 (1907).

\bibitem{Jackson75}
J. D. Jackson, {\it Classical electrodynamics}, 2nd ed., 
(John Willey \& Sons, 1975).

\bibitem{epsilonAg}
{\it Handbook of Optical Constants of
Solids }, edited by E.D. Palik  (Academic, Orlando, 1985).

\bibitem{caution}
While the discussion has general validity, we give approximate 
analytic expressions 
for reference, valid for perfect metals and $d\ll L$.

\end{references}
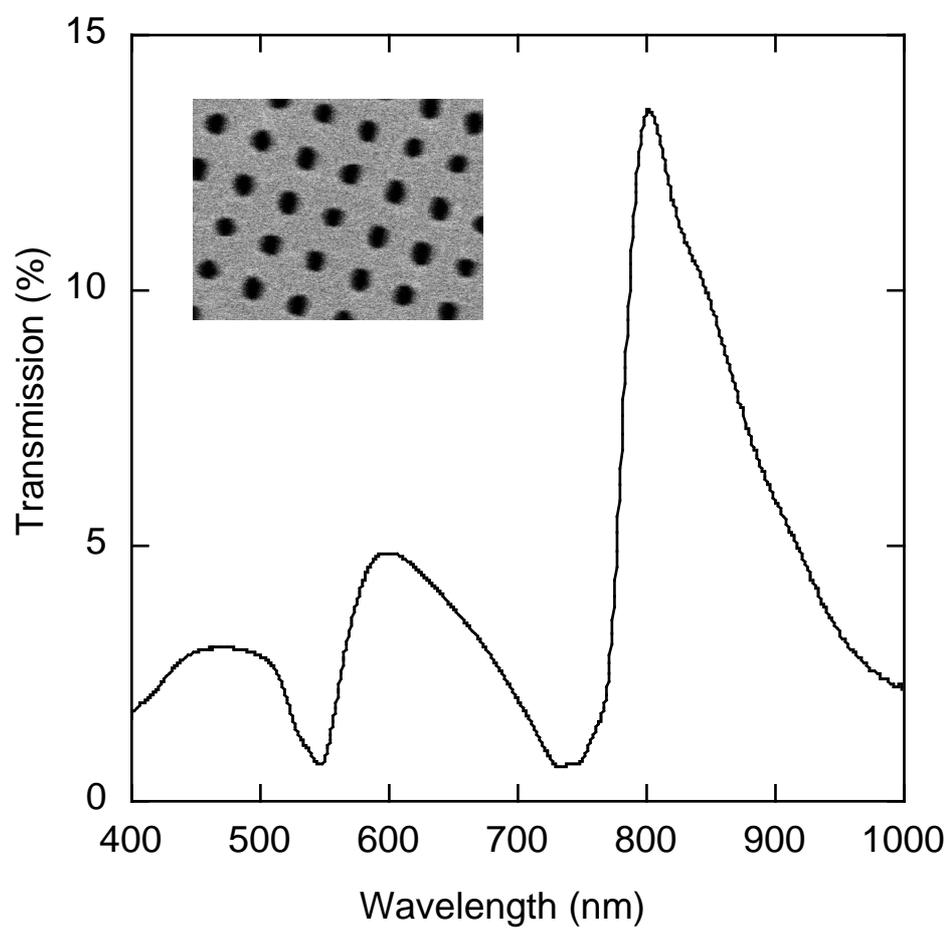
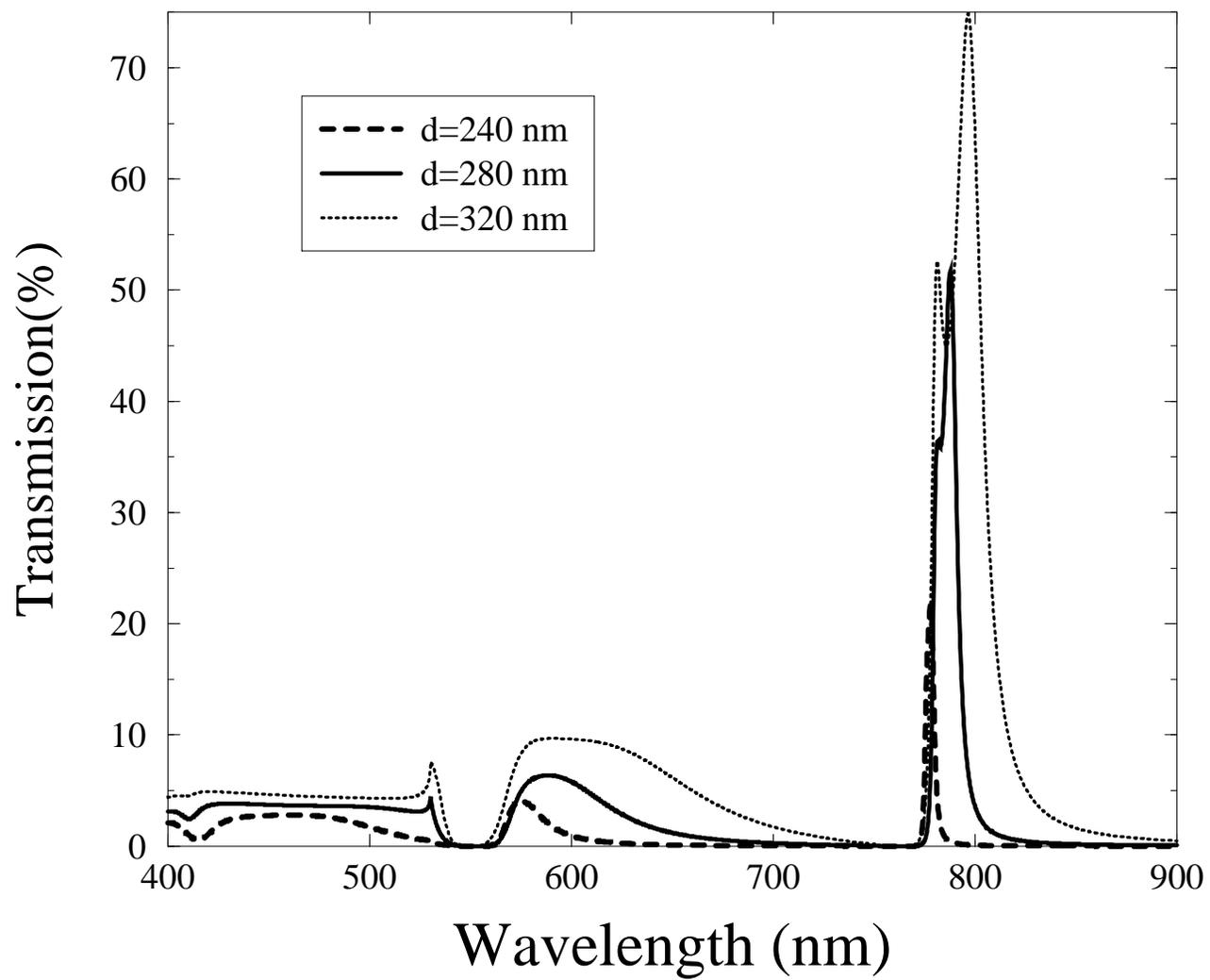
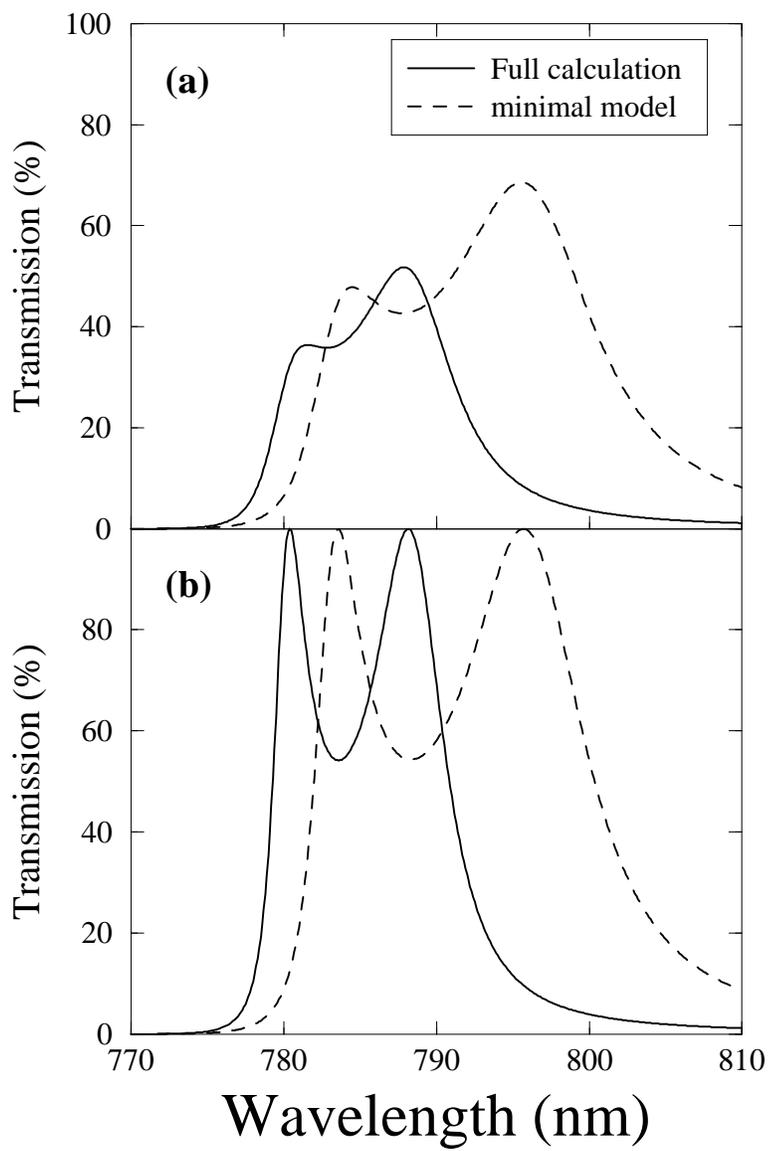
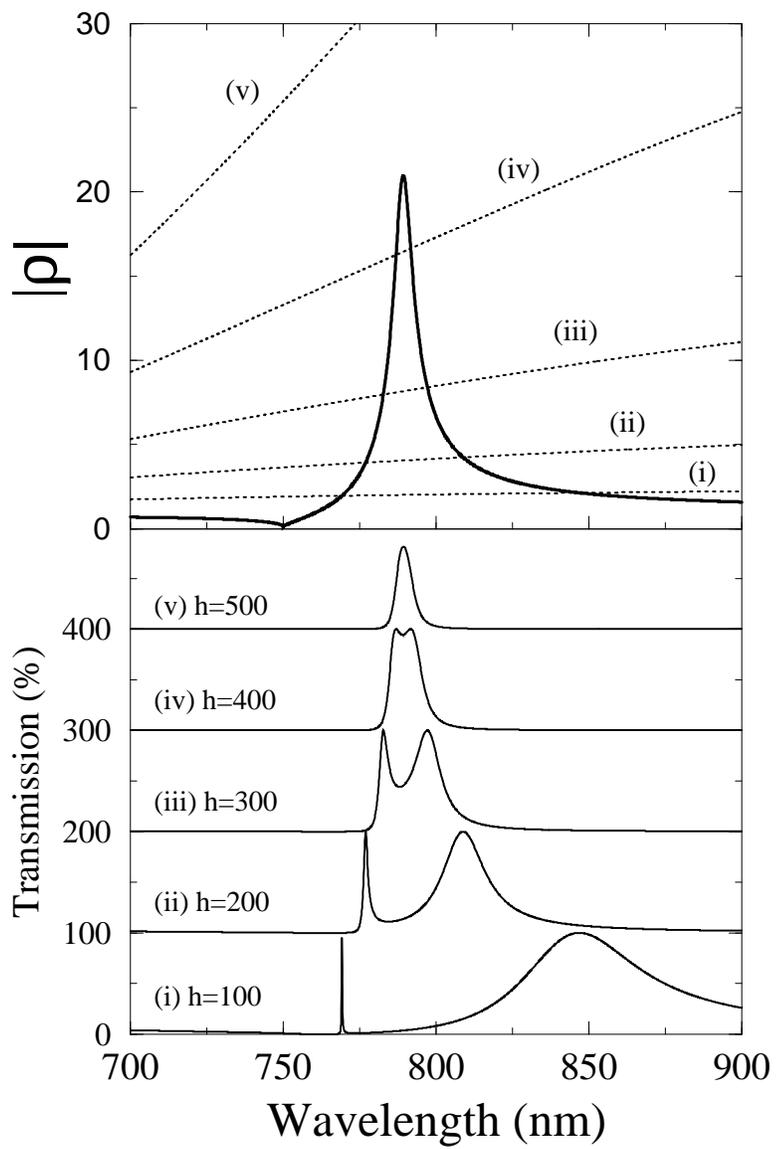
\end{document}